\documentclass[manuscript,appendixfloats,numberedappendix,showkeys]{aastex6}

\usepackage{graphicx}
\usepackage{amsmath}

\usepackage[utf8]{inputenc}
\usepackage[english]{babel}

\fullcollaborationName{The Friends of AASTeX Collaboration}

\begin{document}


\title{A Simple Formula for a Planet's Mean Annual Insolation by Latitude}


\author{Alice Nadeau\altaffilmark{1} and Richard McGehee}
\affil{University of Minnesota \\
School of Mathematics \\
206 Church St. SE\\
Minneapolis, MN 55455}


\altaffiltext{1}{nadea093@umn.edu}

\begin{abstract}
In this paper, we use a sixth order Legendre series expansion to approximate the mean annual insolation by latitude of a planet with obliquity angle $\beta$, leading to faster computations with little loss in the accuracy of results.  We discuss differences between our method and selected computational results for insolation found in the literature.

\end{abstract}


\section{Introduction}

Incoming solar radiation is an important input in many earth systems models.  This physical quantity is needed in areas ranging from low-dimensional energy balance models (e.g. the Budyko energy balance model \citep{Budyko}) to large global circulation models (GCMs), e.g. NASA's ModelE AR5 (available on NASA's ModelE website), or earth systems models (ESMs).  It is common practice to compute insolation by latitude using computer algorithms.  For example, NASA's latitudinal insolation calculations for ModelE AR5 rely on three FORTRAN subroutines that 1) calculate Earth's orbital parameters (eccentricity, obliquity, and longitude of perihelion) as a function of year, 2) calculate distance to the sun and declination angle as functions of time of year and orbital parameters, and 3) calculate the time integrated zenith angle as a function of the declination angle and the time interval of the day.

These computer calculations are useful for models with latitude-longitude grids (as is typical in GCM's or ESM's); however, to convert this information to useable data for other modeling scenarios is not always straightforward.  For example, in the Budyko-Widiasih energy balance model, one must know the annual average insolation as a function of latitude in order to make use of the model \citep{Widiasih}.  Obtaining such a function by fitting a polynomial, a trigonometric function, or a spline to data points given by a computer program obscures the true relationship between insolation and latitude and may introduce errors that, when integrated over time scales of millennia, give meaningless results.

In the following section we give the results of an integration method used in several sources \citep{Dobrovolskis,McGehee2012,Ward} to find the mean annual insolation by latitude for any planet as a function of obliquity and eccentricity and present our results, a sixth-degree approximation to the insolation distribution.  In Section \ref{Section-Planets} we give examples of this approximation to the insolation distributions of Earth, Mars, and Pluto.  We conclude with a discussion of the applicability of these approximations and some interesting mathematical conjectures requiring further investigation.

\section{Mean Annual Insolation Function}
\label{Section-Insolation-Function}

It has been shown in several sources  that one can calculate (as a function of latitude) the mean annual insolation of a swiftly rotating planet using only first principles \citep{Dobrovolskis,McGehee2012,Ward}.  Following the notation in McGehee and Lehman, \citep{McGehee2012}, one can express mean annual insolation $\overline{I}$ as a function of eccentricity $e$, obliquity $\beta$, and sine of latitude $y$ by finding the insolation at any point on the Earth's surface, integrating over the course of one orbital period, then integrating over all longitudes \citep[see][Section~4]{McGehee2012}.  Their results are
\[\overline I(e,y,\beta)=Q(e)s(y,\beta)\]
where the distribution of insolation across the sine of the latitude is given by
\begin{equation}
s(y,\beta)=\frac{2}{\pi^2}\int_{0}^{2\pi}\sqrt{1-\left(\sqrt{1-y^2}\sin\beta\sin\gamma-y\cos\beta\right)^2}\, d\gamma
\label{EQ:distribution}
\end{equation}
(where $\gamma$ is longitude) and the magnitude of insolation is given by
\begin{equation}
Q(e)=\frac{Q_0}{\sqrt{1-e^2}}.
\label{EQ:global}
\end{equation}
where $Q_0$ is the global annual average insolation \citep{McGehee2012}.  We see that their analysis is general enough to apply to any planet orbiting a star with a spin period much shorter than its orbital period, as long as the appropriate physical parameters are known.

In the appendix we present a recipe for approximating the distribution function $s(y,\beta)$ as a polynomial in $y$ and $\beta$ to any desired degree of accuracy.  In Section \ref{Section-Planets} we show that the sixth-degree approximation is sufficient for most purposes, and is given by
\begin{align}
\sigma_{6}(y,\beta)= 1-\frac{5}{8}p_2(\cos\beta)p_2(y)-\frac{9}{64}p_4(\cos\beta)p_4(y)-\frac{65}{1024}p_6(\cos\beta)p_{6}(y)
\label{EQ-Approximation}
\end{align}
where the $p_k$'s are the Legendre polynomials
\begin{align*}
p_2(y)&=(3y^2-1)/2\\
p_4(y)&=(35y^4-30y^2+3)/8\\
p_6(y)&=(231y^6-315y^4+105y^2-5)/16
\end{align*}
It should be noted that the polynomial $\sigma_6(y,\beta)$ is the best least-mean-square approximation to the function $s(y,\beta)$.  There may be better uniform approximations, but that possibility is not explored here.

\citet{North} explicitly gives a second-degree approximation for the insolation distribution for the Earth as
\[\hat\sigma_2(y)=1-.482p_2(y),\]
 stating that the approximation to the actual distribution is accurate to within $2\%$.  North notes that this approximation was first given in \citet{Chylek} as a linear interpolation of the insolation distribution, although no closed-form formula is given in that paper.  Since this approximation was computed only for the current obliquity of the Earth, it cannot be used to compute changes due to the Milankovitch cycles nor can it be used for other planets. We suggest that the polynomial approximation, $\sigma_6$ given above be used instead of the integral form of the insolation distribution function because the approximation is more computationally efficient and sufficiently accurate to capture the qualitative characteristics of the actual distribution function.

\section{Planetary Examples} 
\label{Section-Planets}

\begin{figure}
\begin{center}
\includegraphics[width=.7\textwidth]{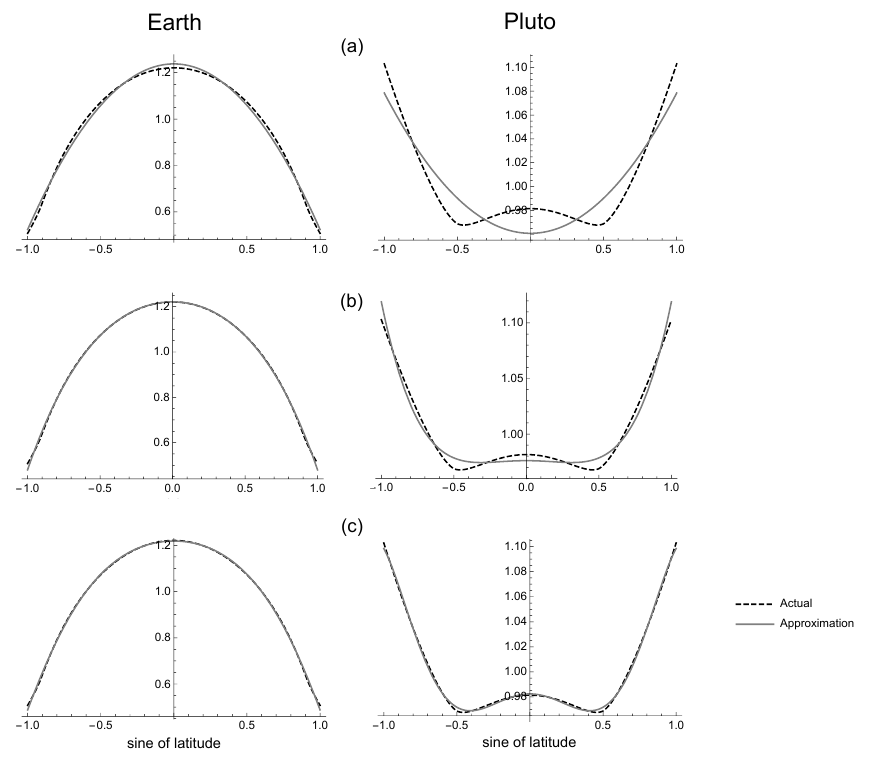}
\end{center}
\caption{Insolation distributions for Earth and Pluto, both actual (dashed) and various approximations (solid) are shown.  In this figure we show (a) second-degree approximations ($\sigma_2$), (b) fourth-degree approximations ($\sigma_4$), and (c) sixth-degree approximations ($\sigma_6$).  It is not until the sixth-degree approximation that we capture the slight `W' shape of Pluto's insolation distribution. }
\label{Insolation-Earth-Pluto}
\end{figure}

The formula for $\sigma_6(y,\beta)$ can be truncated to produce second- and fourth-degree polynomials in $y$ and $\cos\beta$ in the form of
\[\sigma_{2N}(y,\beta)=1+ \sum_{n=1}^N q_{2n}(\beta)p_{2n}(y)\]
for $N=1,2,3.$

As stated in the previous section, North used a second-degree approximation in his analysis of a simple climate model of the Earth.  \cite{McGehee2014} also used North's second-degree approximation in their analysis of the ice-albedo feedback mechanism in a paleoclimate model .  A second-degree approximation is sufficient for Earth's obliquity ($\beta\approx23.4^\circ$); however, other obliquity angles produce qualitatively different insolation distributions, and the second-degree approximation is no longer sufficient to capture an accurate insolation distribution.  

For example, consider Pluto, which has obliquity $\beta=119.6^\circ$.  In Figure \ref{Insolation-Earth-Pluto} we show the insolation distributions of Earth and Pluto plotted with second-, fourth-, and sixth-degree approximations.  Notice that even though the second-degree approximation does very well in matching Earth's insolation distribution, we need at least a sixth-degree approximation to capture the  slight `W' shape of Pluto's insolation distribution.

\begin{figure}
\begin{center}
\includegraphics[width=.95\textwidth]{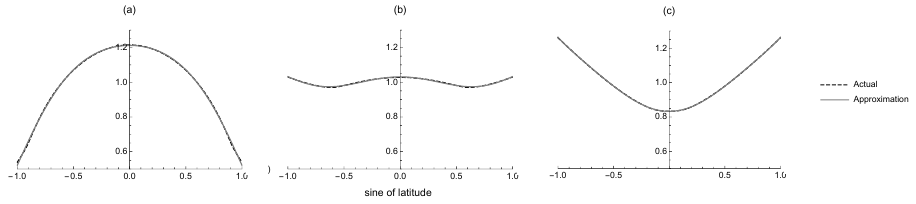}
\end{center}
\caption{Insolation distribution for various obliquity values in Mars' orbital history.  In all plots we give the actual (dashed) and sixth-degree approximation (solid) to the insolation distribution.  In (a) we have $\beta=37.62^\circ$, which is Mars' average obliquity as given in \cite{Laskar}, in (b) we have $\beta=54^\circ$, which gives an example where polar and equatorial regions receive about the same yearly insolation, and in (c) we have $\beta=82.035^\circ$, which is the maximal possible obliquity, \citep{Laskar}. }
\label{Insolation-Mars}
\end{figure}

Being able to compute insolation by latitude as an explicit function of obliquity is particularly important in the case of Mars due to the chaotic nature of Mars' obliquity over the course of 5 billion years. \cite{Laskar} showed that the obliquity of Mars ranges from $\beta\approx24^\circ$ to $\beta\approx82^\circ$.  Over this range in obliquity, the insolation distribution changes drastically, going from a downward facing parabolic shape, to a strong `W' shape, to an upward facing parabolic shape (see Figure \ref{Insolation-Mars}).  In modeling the climate of Mars over time, it would be beneficial to have an algebraic representation of the insolation with explicit dependence on obliquity. 

\section{Discussion}  

In order to understand fully a planet's climatic history or future, one must understand the amount of solar radiation reaching the planet's outer atmosphere.  Understanding how all orbital parameters affect a planet's insolation is crucial for implementing insolation into any sort of model, whether it is a GCM or a low-dimensional dynamical system.  The Legendre approximation to sixth-degree given in this paper gives a good approximation for most obliquities, and if more accuracy is desired or needed, one may simply use equations \ref{s2N-eq} and \ref{c2k-eq} to compute an approximation to the desired accuracy.

\acknowledgements

\section{Acknowledgements}

The authors gratefully acknowledge the support of the Mathematics and Climate Research Network (NSF Grants DMS-0940366 and DMS-0940363).

\appendix

 \section{Computing the Approximation} 

 Although equation (\ref{EQ:distribution}) gives the insolation for rapidly rotating planets, using the integral formula in low-dimensional dynamical systems models of the earth is cumbersome.  To bypass this issue, several sources instead fix $\beta$ and use a second-degree Legendre approximation in $y$ to compute mean annual insolation by latitude \citep{Chylek, McGehee2012, McGehee2014, North,Widiasih}.  A second-degree approximation is sufficient in the case of Earth but lacking in the case of other planets due to qualitative changes in the insolation distribution as $\beta$ changes.  In the following section we compute the approximation up to arbitrary degree and explicitly give the Legendre approximation to sixth-degree as a function of both $y$ and $\beta$.
 
Because the Legendre polynomials, $p_n$, are orthogonal functions we may write the insolation distribution function as a Legendre series in $y$:
\begin{align*}
s(y,\beta)=\sum_{n=0}^\infty q_{2n}(\beta)p_{2n}(y)
\end{align*} 
where $q_{2n}$ is to be determined. Notice that we only take the even Legendre polynomials because $s$ is an even function of $y$ \citep{Dobrovolskis}.

We want to compute approximations, $\sigma_{2N}(y,\beta) \approx s(y,\beta)$, of the insolation function of the form
\begin{align}
\sigma_{2N}(y,\beta)= 1+q_2(\beta)p_2(y)+q_4(\beta)p_4(y)+\cdots+q_{2N}(\beta)p_{2N}(y).
\label{Insolation-approximation}
\end{align}
The orthogonality of the Legendre polynomials allows us to say
\[q_{2n}(\beta)=\frac{\int_{0}^{1}p_{2n}(y)s(y,\beta)dy}{\int_{0}^{1}p_{2n}^2(y)dy}.\] 
Let
\begin{align*}
c_{2k}(\beta)&=\frac{2}{\pi^2}\int_{0}^{1}y^{2k}\,s(y,\beta) dy\\
&=\frac{2}{\pi^2}\int_{0}^{1}\int_{0}^{2\pi}y^{2k}\,\sqrt{1-(\sin\beta\sqrt{1-y^2}\cos\theta-y\cos\beta)^2}\,d\theta dy\\
&=\frac{1}{\pi^2}\int_{-\pi/2}^{\pi/2}\int_{0}^{2\pi}\sin^{2k}\phi\,\sqrt{1-(\sin\beta\cos\phi\cos\theta-\cos\beta\sin\phi)^2}\cos\phi\,d\theta d\phi
\end{align*}
for any nonnegative $k$ where $\phi$ is the latitude. If we write the even-degree Legenrdre polynomials as
\[p_{2n}(y)=\sum_{k=0}^na_{2k}y^{2k}\]
then  
\begin{align}
q_{2n}(\beta)=P_{2n}\sum_{k=0}^na_{2k}c_{2k}(\beta).
\label{s2N-eq}
\end{align}
 where 
 \[\frac{1}{P_{2n}}={\int_{0}^{1}p_{2n}^2(y)dy}.\]
Now we must calculate the $c_{2k}$'s.  We will show that 
\begin{align}
c_{2k}(\beta)=\sum_{\ell=0}^{k}{2k \choose 2\ell}(\sin\beta)^{2(k-\ell)}(\cos\beta)^{2\ell}\frac{1}{\pi^2}\left(\int_{-\pi/2}^{\pi/2} (\cos\hat\phi)^{2(k+1-\ell)}(\sin\hat\phi)^{2\ell}     \, d\hat\phi\right) \left(\int_{0}^{2\pi}(\cos\hat\theta)^{2(k-\ell)} \,d\hat \theta \right).
\label{c2k-eq}
\end{align}

We can change the representation of $c_{2k}(\beta)$ from a double integral into a triple integral over the unit sphere,
\begin{align*}
c_{2k}(\beta)&=\frac{1}{\pi^2}\int_{-\pi/2}^{\pi/2}\int_{0}^{2\pi}\sin^{2k}\phi\,\sqrt{1-(\sin\beta\cos\phi\cos\theta-\cos\beta\sin\phi)^2}\cos\phi\,d\theta d\phi\\
&=\frac{3}{\pi^2}\int_{0}^{1}\int_{-\pi/2}^{\pi/2}\int_{0}^{2\pi}\sin^{2k}\phi\,\sqrt{1-(\sin\beta\cos\phi\cos\theta-\cos\beta\sin\phi)^2}\,r^2\cos\phi\,d\theta d\phi dr,
\end{align*}
because 
\[\int_{0}^1r^2dr=\frac{1}{3}.\]
This leaves us with a volume integral with the region of integration being the unit sphere.  Let's change variables back to cartesian coordinates with the change of variables
\begin{align*}
x&=r\cos\phi\cos\theta\\
y&=r\cos\phi\sin\theta\\
z&=r\sin\phi.
\end{align*}
Implementing this change of variables in the above integral gives us
\begin{align*}
c_{2k}(\beta)&=\frac{3}{\pi^2}\iiint_{S^2}\left(\frac{z}{r}\right)^{2k}\sqrt{1-(\sin\beta (x/r)-\cos\beta (z/r))^2} \,dV\\
&=\frac{3}{\pi^2}\iiint_{S^2}\left(\frac{z}{r}\right)^{2k}\frac{1}{r}\sqrt{r^2-(x\sin\beta -z\cos\beta )^2} \,dV\\
&=\frac{3}{\pi^2}\iiint_{S^2}\left(\frac{z^2}{x^2+y^2+z^2}\right)^{k}\frac{1}{\sqrt{x^2+y^2+z^2}}\sqrt{(x^2+y^2+z^2)-(x\sin\beta -z\cos\beta )^2} \,dV.
\end{align*}

We'd like to get the $\sin\beta$ and $\cos\beta$ out from under the radical, so we implement the rotational change of variables
\begin{align*}
x&=\hat x\cos\beta-\hat z\sin\beta\\
y&=\hat y\\
z&=\hat x\sin\beta+\hat z\cos\beta.
\end{align*}
We see that the determinant of the Jacobian is $1$ and that
\[x^2+y^2+z^2=\hat x^2+\hat y^2+\hat z^2,\]
so changing variables gives us
\begin{align*}
c_{2k}(\beta)&=\frac{3}{\pi^2}\iiint_{S^2}\left(\frac{(\hat x\sin\beta+\hat z\cos\beta)^2}{\hat x^2+\hat y^2+\hat z^2}\right)^{k}\frac{1}{\sqrt{\hat x^2+\hat y^2+\hat z^2}}\sqrt{(\hat x^2+\hat y^2+\hat z^2)-\hat z^2} \,d\hat V\\
&=\frac{3}{\pi^2}\iiint_{S^2}\left(\frac{(\hat x\sin\beta+\hat z\cos\beta)^2}{\hat x^2+\hat y^2+\hat z^2}\right)^{k}\frac{1}{\sqrt{\hat x^2+\hat y^2+\hat z^2}}\sqrt{\hat x^2+\hat y^2} \,d\hat V
\end{align*}
Notice that we still have a volume integral over the unit sphere because the rotation doesn't change the region of integration.  Let's return to spherical coordinates with the change of variables
\begin{align*}
\hat x&=\hat r\cos\hat\phi\cos\hat\theta\\
\hat y&=\hat r\cos\hat\phi\sin\hat\theta\\
\hat z&=\hat r\sin\hat\phi.
\end{align*}
Here the determinant of the Jacobian is $\hat r^2\cos\hat\phi$ so the integral becomes
\begin{align*}
c_{2k}(\beta)&=\frac{3}{\pi^2}\int_{0}^{1}\int_{-\pi/2}^{\pi/2}\int_{0}^{2\pi}\left( \cos\hat\phi\cos\hat\theta\sin\beta+\sin\hat\phi\cos\beta\right)^{2k} \hat r^2\cos^2\hat\phi\,d\hat \theta d\hat\phi d\hat r\\
&=\frac{1}{\pi^2}\int_{-\pi/2}^{\pi/2}\int_{0}^{2\pi}\left( \cos\hat\phi\cos\hat\theta\sin\beta+\sin\hat\phi\cos\beta\right)^{2k} \cos^2\hat\phi\,d\hat \theta d\hat\phi \\
&=\frac{1}{\pi^2}\int_{-\pi/2}^{\pi/2}\int_{0}^{2\pi}\left( \sum_{\ell=0}^{2k}{2k \choose \ell}(\cos\hat\phi\cos\hat\theta\sin\beta)^{2k-\ell}(\sin\hat\phi\cos\beta)^\ell\right) \cos^2\hat\phi\,d\hat \theta d\hat\phi \\
&=\sum_{\ell=0}^{2k}{2k \choose \ell}(\sin\beta)^{2k-\ell}(\cos\beta)^{\ell}\frac{1}{\pi^2}\left(\int_{-\pi/2}^{\pi/2} (\cos\hat\phi)^{2(k+1)-\ell}(\sin\hat\phi)^\ell     \, d\hat\phi\right) \left(\int_{0}^{2\pi}(\cos\hat\theta)^{2k-\ell} \,d\hat \theta \right)\\
\end{align*}

Since the insolation function is an even function with respect to $\beta$ \citep{Dobrovolskis}, we must have a zero term for all $\ell$ odd in the above formula.  This zero term is easy to see from the first integral in the final line above.  Notice that for odd $\ell$ we have an odd function over a symmetric region centered at 0, meaning the integral is zero, and thus that the entire term is zero.  Then we have
\begin{align*}
c_{2k}(\beta)&=\sum_{\ell=0}^{k}{2k \choose 2\ell}(\sin\beta)^{2(k-\ell)}(\cos\beta)^{2\ell}\frac{1}{\pi^2}\left(\int_{-\pi/2}^{\pi/2} (\cos\hat\phi)^{2(k+1-\ell)}(\sin\hat\phi)^{2\ell}     \, d\hat\phi\right) \left(\int_{0}^{2\pi}(\cos\hat\theta)^{2(k-\ell)} \,d\hat \theta \right)
\end{align*}
as equation \ref{c2k-eq} indicated.

Now we may begin calculating the $c_{2k}$'s to find the $q_{2n}$'s.  The results follow from simple calculus computations, so only the results are shown below.  The $c_{2k}$'s to sixth-degree in $\sin\beta$ are:
\begin{align}
c_{2}(\beta)&=\frac{1}{8}(-\cos^2\beta-3)\label{c2}\\
c_{4}(\beta)&=-\frac{1}{64}(\cos^4\beta+6\cos^2\beta-15)\label{c4}\\
c_6(\beta)&=\frac{5}{1024}(-\cos^6\beta-3\cos^4\beta-15\cos^2\beta+35)\label{c6}.
\end{align}
Then we may calculate the $q_{2N}$'s up to $N=3$ by plugging the above functions into equation \ref{s2N-eq} to get:
\begin{align}
q_{2}(\beta)&=-\frac{5}{16}(3\cos^2\beta-1)=-\frac{5}{8}p_2(\cos\beta)\label{s2}\\
q_{4}(\beta)&=-\frac{9}{512}\left(35\cos^4\beta-30\cos^2\beta+3\right)=-\frac{9}{64}p_4(\cos\beta) \label{s4}\\
q_6(\beta)&=-\frac{65}{16384}\left(231\cos^6\beta - 315\cos^4\beta + 105\cos^2\beta - 5\right)=-\frac{65}{1024}p_6(\cos\beta).\label{s6}
\end{align}

%
%
%
%

\allauthors

\listofchanges

\end{document}